\documentclass{article}
\usepackage{amsmath, amsfonts, amssymb, bm, graphicx, newtxmath, float, authblk}
\newcommand{\probP}{\text{I\kern-0.15em P}}
\newcommand{\Cov}{\mathrm{Cov}}
\title{A novel statistical framework for the analysis of the degree of technology adoption}
\author[1]{Vahidin Jeleskovic}
\author[1]{David Alexander Behrens}
\author[1 2 3 4 5]{Wolfgang Karl Härdle}
\affil[1]{School of Business and Economics, Humboldt University of Berlin, Unter den Linden 6, 10099, Berlin, Germany}
\affil[2]{Blockchain Research Center
Ladislaus von Bortkiewicz Professor
Unter den Linden 6,  10099 Berlin, Germany}
\affil[3]{Xiamen University - Wang Yanan Institute for Studies in Economics (WISE)
A 307, Economics Building
Xiamen, Fujian 10246
China}
\affil[4]{Charles University
Professor
Celetná 13
Dept Math Physics
Praha 1, 116 36
Czech Republic}
\affil[5]{National Yang Ming Chiao Tung University
No. 1001, Daxue Rd. East Dist.
Hsinchu City 300093
Taiwan}

\begin{document}
\maketitle
\begin{abstract}
Technology adoption research aims to determine the reasons why and how individuals, corporations, and industries start using new technology. Furthermore, technology adoption itself is decomposed into underlying sub-processes which are characterized by a finite number of sequential states in order to capture its evolutionary nature. Building upon that, in this paper a technology adoption index is being constructed that allows for statistical testing. This new framework is flexible with respect to the number of underlying models, and accounts for nonlinearities within the evolution of technology adoption. It can be considered as novel because it gives opportunity to a quantitative analysis of technology adoption that has not existed before. Subsequently, this framework is applied for an integrated model of technology adoption. 
\end{abstract}
\section{Introduction}
Over the past decades, new technologies have been developed at a fast pace. Some can be considered as being highly successful, like the internet, the personal computer, or smartphones, whereas others failed to be adopted and, therefore, remain unknown to the vast majority of people. The research on technology adoption aims to identify its main determinants and to decompose it into clearly distinguishable steps towards a full adoption.\\
The first breakthroughs in this field occurred in the second half of the twentieth century. While early models were concentrated on the acceptance of technology and psychological modeling of individual usage behavior, later models focused on the evolutionary nature of the process itself, namely the maturity of technology usage. In recent years, researchers shifted towards combining existing models in order to capture different aspects of technology adoption jointly.\\
Although technology adoption research can be considered to be a well-established field, it is mostly restricted to qualitative or descriptive empirical research. With this paper, we aim to add to the existing research by proposing a novel framework that allows for quantitative analyses and statistical testing. It is supposed to provide the tools to compare the degree of technology adoption of different industries, and to test differences in technology adoption between single corporations and the industries they are located in, between different industries, and also between different technologies. In order to do so, the paper is organized as follows:\\
First, we will start with a literature overview, summarizing major publications regarding technology adoption research. Second, the novel Integrated Technology Adoption Index is introduced. In this section, the formal framework for the construction of the index is provided, followed by the introduction of the index in its linear and nonlinear form. Third, a statistical testing procedure is pointed out. In order to demonstrate its validity, the statistical properties and asymptotics of the Integrated Technology Adoption Index are being examined, and, subsequently, used to form a test statistic. Finally, the quantitative framework is applied to an integrated model of technology adoption.
\section{Literature Overview}
Early theories and models concerning technology acceptance from a decision-making perspective are linked to the Theory of Reasoned Action (TRA) (Fishbein \& Ajzen, 1975; Ajzen \& Fishbein, 1980). TRA is a theory located in the field of cognitive psychology, aiming to explain individual behavior in specific situations. The factors used to predict an individual's behavior are behavioral beliefs about the outcome of a certain action that form its attitude towards that behavior, subjective norms that account for the perception of others, and, finally, the actual intention to show a certain behavior (LaCaille, 2020). Although it was not originally developed for understanding technology acceptance, it yet had a meaningful impact on subsequent literature. A theory that can be seen as a relative of TRA is the Theory of Planned Behavior (TPB) (Ajzen, 1985, 1991). TPB is highly similar to TRA but includes the ability to control as a behavioral predictor, and the individual's attitude is decomposed into perceived usefulness, perceived ease of use, and compatibility.\\
The Technology Acceptance Model (TAM) was designed in order to understand the user acceptance of novel developments in information systems (Davis, 1985; Davis et al. 1989). Here, the actual system use is determined by the behavioral intention to use that is mainly dependent on the individual's attitude towards usage. The two factors that form this attitude are perceived usefulness and perceived ease of use (Davis, 1985). TAM and its extensions are not limited to theoretical modeling and have been used in empirical studies (Szajna, 1996). Further examples are Chau (1996) who examines the usage of office software packages among administrative employees, and Hu et al. (1999) who examine physicians' willingness to engage in telemedicine with its accompanied technologies. Moreover, it is also applied in the context of more recent technology as in Sagnier et al. (2020).\\
In addition, the Perceived Characteristics of Innovating (PCI) model has to be mentioned. It shifts the focus on the individual's perceptions with respect to the adoption of a technology (Moore \&  Benbasat, 1991). Extending ideas of Rogers' Diffusion of Innovation Theory (see Rogers, 1962, 2003), it accounts for the relative advantage, compatibility, complexity, observability, and trialability. Therefore, it takes into consideration that individuals that differ in their cultural background, may behave differently in their adoption of a technology due to their perceptions of it (Poong \& Eze, 2008).\\
Another class of models are maturity models. Their objective is to focus on the evolutionary nature of technology adoption. The most famous representative is the Capability Maturity Model (CMM) (Paulk et al., 1991, 1993). In this class of models, technology adoption is treated as a dynamic process whereby the aim is to divide it into distinguishable stages. The CMM includes five stages: initial, repeatable, defined, managed, and optimizing (Paulk et al. 1993). Both TAM and CMM were extended in subsequent versions (Venkatesh \& Davis, 2000; Venkatesh \& Bala 2008; Paulk et al. 1996; Paulk 1997). Of course, also the CMM was subject to empirical studies, as in Herbsleb et al. (1997) and Garzás \& Paulk (2013).\\
Furthermore, researchers combined different models to integrated models. Examples are Lee et al. (2011) and Lou \& Li (2017) who examined technology adoption by a combination of TAM and Diffusion of Innovation Theory (see Rogers, 1962). A most recent example is Drljevic et al. (2022) who developed an integrated model of TAM and CMM to investigate technology adoption in the context of Blockchain-driven business innovations. It maps two core aspects of technology adoption, novelty and complexity, to acceptance and maturity, respectively. Due to this innovative framework, we choose it as a fruitful example for an application of the statistical framework developed in this paper. 
\section{Integrated Technology Adoption Index}
The first step in a quantitative analysis of technology adoption is to construct an index. This index needs to be flexible in nature with respect to the underlying models used, and straightforward to interpret. Flexibility is obtained by independence of specific models although the focus is on those measuring aspects of technology adoption in a finite sequence of ordered levels. However, interpretability is ensured by normalization.\\
First, the formalism for the construction of such an index is being introduced. Second, a linear version is being developed. Third, the Linear Integrated Technology Adoption Index is extended to a spectrum of nonlinear versions.
\subsection{Formal Framework}
Let there be an arbitrary industry consisting of $n$ corporations, where a corporation is indicated by $i=1,2,\dots, n$, and $k$ underlying models to measure the state of technology adoption, where a model is indicated by $j=1,2, \dots, k$. We will always assume $n>k$. A model $j$ consists of a number of sequential stages which indicate the state of technology adoption in an increasing order. At this point, a distinction has to be made. If the lowest stage of model $j$ indicates no adoption of a certain technology at all, no adjustment is needed. However, if this is not the case, this stage has to be added. Henceforth, a model $j$ that is able to capture the degree of technology adoption with a stage indicating that the technology has not been adopted at all, consists of $m_j +1$ stages. Thus, the random variable $X_{j}\equiv\{a_{j}, p_{a_{j}}; a_{j}=0,1,\dots , m_{j}\}$ is a monotone transformation and takes integer values that correspond to the levels of model $j$, whereby $p_{a_{j}}\equiv\probP (a_{j})$ (Cox, 2012). Since $X_j$ is a discrete random variable, its expectation is given by $\mbox{\sf E}[X_j]=\sum_{a_j=0}^{m_j} a_j p_{a_j}$(Facchinetti \& Osmetti, 2018).\\
In this setting, the following notation is used: $S_j \equiv \mbox{\sf E}[X_j]$. Additionally, it is assumed, given a sample of size $n$, that $r_{a_j}=\sum_{i=1}^n\vmathbb{1}_{x_{ij}=a_j}$. Hence, the Maximum-Likelihood estimate (MLE) of $p_{a_j}$ is $\hat{p}_{a_j}=\frac{r_{a_j}}{n}$, which leads to the following estimate of $S_j$: 
\begin{equation}
\begin{aligned}
\hat{S}_j & = \sum_{a_j=0}^{m_j} a_j \frac{r_{a_j}}{n}\\
& = n^{-1}\sum_{i=1}^n x_{ij}
\end{aligned}
\end{equation}
.\\
The latter expression is convenient because it enables the use of simple matrix algebra. A set of observations is given by\\
\begin{equation*}
\bm{X}= \begin{pmatrix} 
    x_{11} & x_{12} & \dots & x_{1k}\\
    \vdots & \vdots & \ddots &   \vdots \\
    x_{n1} &  x_{n2} &  \dots     & x_{nk} 
    \end{pmatrix}\\
\end{equation*}\\
An element $x_{ij}$ corresponds therefore to a realization of $X_{j}$ at observation $i$.\\
The first step in the construction of an index is to obtain the sample scores $\hat{S}_{j}$ for each model $j$.\\
\begin{equation}
\begin{aligned}
\hat{\bm{S}}&=n^{-1}\bm{1}_{n}^\top \bm{X}\\
&=\begin{pmatrix}
n^{-1}\sum_{i=1}^n x_{i1}, & \dots, & n^{-1}\sum_{i=1}^n x_{ik}
\end{pmatrix}\\
&=\begin{pmatrix}
\hat{S}_{1}, & \dots, & \hat{S}_{k}
\end{pmatrix}
\end{aligned}
\end{equation}\\
Naturally, each $\hat{S}_{j} \in [0,m_{j}]$.\\
The Integrated Technology Adoption Index $I$ (henceforth also called \textit{global} index) is the weighted average of the sub-indices $I_j$. These sub-indices are obtained by a normalization of the scores $S_{j}$ (and the estimates $\hat{I}_j$ from $\hat{S}_j$ accordingly). The weight $w_{j}$ for each sub-index is equal, namely $w_{j}=k^{-1}, \ j=1,2,...,k$. Of course it is possible to choose different weights as long it is ensured that $\sum_{j=1}^{k} w_{j}=1$. However, these cases are not covered here in order to preserve simplicity in the introduction of this framework.\\
Formally, $I: S_1 \times ... \times S_k \to [0,1]$.
\subsection{Linear Index}
The construction of the linear index is obtained by considering a normalization vector $\bm{v}=(\frac{1}{m_{1}},...,\frac{1}{m_{k}})^\top$. The normalization of a score $S_{j}$ yields the sub-index ${I_j}$, with $I_{j} \in [0,1]$.
\begin{equation}
\begin{aligned}
I& \equiv k^{-1}\sum_{j=1}^{k} I_{j}\\
&=k^{-1}\sum_{j=1}^{k} \frac{S_{j}}{m_{j}}\\
&= k^{-1}\bm{S}\bm{v}
\end{aligned}
\end{equation}\\
For the global index $I$ it is also valid that $I \in [0,1]$, which ensures that it is straightforward to interpret, whereby a value of $0$ corresponds to the lowest degree of technology adoption, and a value of $1$ to the highest possible, respectively. Multiplication by a hundred yields the percentage degree of technology adoption. The estimate is easily obtained by
\begin{equation*}
\begin{aligned}
\hat{I}& = k^{-1}\sum_{j=1}^{k} \hat{I}_{j}\\
&=k^{-1}\sum_{j=1}^{k} \frac{\hat{S}_{j}}{m_{j}}\\
&= k^{-1}\hat{\bm{S}}\bm{v}
\end{aligned}
\end{equation*}
All possible values of $I$ form a (hyper-)plane in a $(k+1)$-dimensional space.
\subsection{Nonlinear Index}
The motivation for a nonlinear index is that not every step from one stage to another has the same impact on the degree of technology adoption. A famous analogy is the diffusion of innovation characterized by an S-shaped curve (Rogers, 1962, 2003), or the famous product life cycle (Vernon, 1966).  For the nonlinear index, a greater variety of shapes is allowed.\\
Since the global index is defined as a weighted average, nonlinearities are captured by a different construction of the sub-indices. The main component is the relative difference to the maximum value, given by $\frac{m_j-S_j}{S_j} \in [0,\infty)$. The objective is now to map the relative difference onto the interval $[0,1]$. For this purpose, we make use of the inverse logit transformation (Tomarchio \& Punzo, 2019).
\begin{equation*}
g(y)=\frac{1}{1+\exp(-y)}
\end{equation*}
If this functional form is assumed, extended by a parameter $\alpha_j$, and $y=-\beta_j\log\left(\frac{m_j-S_j}{S_j}\right)$, then the nonlinear sub-index $I_j$ is given by
\begin{equation}
I_{j}=\frac{1}{1+\alpha_j \left(\frac{m_{j}-S_j}{S_j}\right)^{\beta_j}}, \ \ j=1,2,\dots, k
\end{equation}
whereby $\alpha_j >0$ and $\beta_j \geq1$ are shape parameters that enable to account for the individual interpretation of each underlying model. An alternative (numerically stable) representation is $I_{j}=\frac{S_j^{\beta_j}}{S_j^{\beta_j}+\alpha_j \left(m_{j}-S_j\right)^{\beta_j}}$. Of course, other choices could have been made that equally satisfy the formulated requirement. Nonetheless, the nonlinear (sub-)index is defined as above because now the linear version is a special case of its nonlinear extension; that is when $\alpha_j=\beta_j=1$. 
\begin{equation*}
\begin{aligned}
I_{j | \alpha_j=1, \beta_j=1}&=\frac{1}{1+\frac{m_j-S_j}{S_j}}\\
&=\frac{S_j}{m_j}
\end{aligned}
\end{equation*}\\
While an increasing $\alpha_j$ dampens how fast $I_j$ grows in $S_j$, an increase in $\beta_j$ leads to a higher steepness in the middle part of the curve. This setup allows for concave, convex, s-shaped, and linearly shaped indices. The usage of $\hat{S}_j$ yields again the estimate $\hat{I}$.\\
\begin{figure}[H]
 \centering
\includegraphics[scale=0.5]{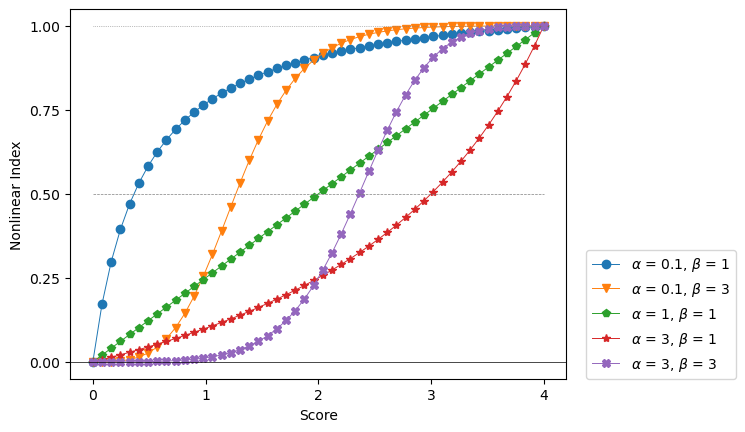}
\caption{Plot of nonlinear sub-indices with different $\alpha$ and $\beta$ against a score with $m=4$. }
\end{figure}
\section{Statistical Testing}
The ultimate objective of this quantitative approach is to enable statistical testing. Two important questions to examine could be whether a single corporation is significantly different in its degree of adoption of a certain technology when compared to its whole industry, or if two distinct industries differ from each other in the adoption of a certain technology. The first step is to clarify the statistical properties and asymptotics of the technology adoption index and its components. Subsequently, test statistics can be formed.
\subsection{Statistical Properties and Asymptotics}
In Facchinetti \& Osmetti (2018) it is pointed out that $r_{a_j}$ is a binomial random variable with $\mbox{\sf E}[r_{a_j}] =n p_{a_j}$. Thus,
\begin{equation*}
\begin{aligned}
\mbox{\sf E}[\hat{S}_j]&=\sum_{a_j=0}^{m_j} a_j \frac{\mbox{\sf E}[r_{a_j}]}{n}\\
&=\sum_{a_j = 0}^{m_j} a_j p_{a_j}
=S_j
\end{aligned}
\end{equation*}
which ensures unbiasedness of the estimator. Furthermore, it is a consistent estimator (Facchinetti \& Osmetti, 2018). This can also be easily shown under the assumption that the $x_{ij}$, $i=1,2,...,n$ are identically, independently distributed (iid) with mean $S_j$ and variance $\mathbb{V}[x_{ij}]\equiv \sigma^2_j$. The variance of the estimator $\hat{S}_j$ is given by
\begin{equation*}
\begin{aligned}
\mathbb{V}[\hat{S}_j]&=\mathbb{V}\left[\sum_{a_j=0}^{m_j} a_j \frac{r_{a_j}}{n}\right]\\
&=\mathbb{V}\left[n^{-1}\sum_{i=1}^{n}x_{ij}\right]\\
&=n^{-2}\sum_{i=1}^{n}\sigma_j^2\\
&=\frac{\sigma^2_j}{n}
\end{aligned}
\end{equation*}
It is obvious that $\mathbb{V}[\hat{S}_j] \rightarrow 0$ as $n\to\infty$.
Now shifting the focus to the asymptotics, we will make use of Central Limit Theorem:
\begin{equation*}
\sqrt{n}\frac{\hat{S}_j-S_j}{\sigma_j}\overset{\cal{L}} = Z, \ n \rightarrow \infty
\end{equation*}
where $Z \sim \mathcal{N}(0,1)$. Thus: $\hat{S}_j \overset{\cal{L}}{\longrightarrow} \mathcal{N}(S_j, \sigma^2_j/n)$ as $n\rightarrow \infty$ (see Härdle \& Simar, 2015).\\
From this point, it is obvious to see that the asymptotics of the linear and nonlinear version differ from each other. Since the asymptotic distribution of the linear index is straightforward, we will start with it.
\begin{align}
\hat{I}_j = \frac{\hat{S}_j}{m_j} \overset{\cal{L}}{\longrightarrow} \mathcal{N}(I_{j},\sigma^{2}_j/ n m^{2}_j)
\end{align}\\
Hence, the sum of sub-indices (and their estimates) is also normally distributed. Since the different factors $j$ in terms of different models for technology adoption  cannot be assumed to be independent, one has to account for this fact in the asymptotic distribution. From now on, let $\Sigma_j=\sigma^{2}_j/ n m^{2}_j$. Thus:
\begin{align}
\sum_{j=1}^k \hat{I}_j \overset{\cal{L}}{\longrightarrow} \mathcal{N}\left(\sum_{j=1}^k I_j, \sum_{j=1}^k \Sigma_j + 2\sum_{j \neq \tilde{j}}\rho_{j\tilde{j}}\sqrt{\Sigma_j \Sigma_{\tilde{j}}}\right)
\end{align}
and
\begin{align}
\hat{I}\overset{\cal{L}}{\longrightarrow} \mathcal{N}\left(I, k^{-2}\left\{\sum_{j=1}^k \Sigma_j + 2\sum_{j \neq \tilde{j}}\rho_{j\tilde{j}}\sqrt{\Sigma_j \Sigma_{\tilde{j}}}\right\}\right)
\end{align}
with correlation coefficient $\rho_{j\tilde{j}}$ that measures correlation between $\hat{I}_j$ and $\hat{I}_{\tilde{j}}$, where $j$ and $\tilde{j}$ indicate two different models such that $j \neq \tilde{j}$.\\
For the derivation of the asymptotic distribution of the nonlinear index, we will make use of the well-known Delta Method. According to it, if there is a parameter $\theta$ and a real valued function $f(t)$, $f:\mathbb{R}\to\mathbb{R}$, that is differentiable at $\theta$, and if 
\begin{align*}
\sqrt{n}(\hat{\theta}-\theta)\overset{\cal{L}}{\longrightarrow}\mathcal{N}(0,\sigma^2),\ n\rightarrow \infty
\end{align*}
then it holds that
\begin{align*}
\sqrt{n}(f(\hat{\theta})-f(\theta))\overset{\cal{L}}{\longrightarrow}\mathcal{N}(0,\sigma^2D^2),\ n\rightarrow \infty
\end{align*}
with $D=\frac{\partial f}{\partial t}\Bigr|_{\substack{t=\theta}}$ (see Härdle \& Simar, 2015).
Now, let $f_j(t)$ correspond to the functional form of the nonlinear index $I_j$
\begin{equation*}
f_j(t)=\frac{1}{1+\alpha_j \left(\frac{m_{j}-t}{t}\right)^{\beta_j}}
\end{equation*}
such that
\begin{equation}
D_j=\frac{\partial f_j(t)}{\partial t}\Bigr|_{\substack{t=S_j}}=\frac{\alpha_j \beta_j m_j \left(\frac{m_j-S_j}{S_j}\right)^{\beta_j}}{\left(1+\alpha_j\left(\frac{m_j-S_j}{S_j}\right)^{\beta_j}\right)^2 (m_j-S_j)S_j}
\end{equation}
According to the (uni-variate) Delta Method, the asymptotic distribution is given by
\begin{align}
\sqrt{n}(\hat{I}_j-I_j)\overset{\cal{L}}{\longrightarrow}\mathcal{N}(0,\sigma^2_j D^2_j/m^2_j)
\end{align}
if $D_j$ exists, which is valid for $S_j \in (0,m_j)$ (meaning that there is a variance greater than zero within the population), and, therefore,
\begin{align}
\hat{I} \overset{\cal{L}}{\longrightarrow} \mathcal{N}\left(I, k^{-2}\left\{\sum_{j=1}^k \Sigma_jD_j^2+2\sum_{j\neq\tilde{j}}\rho_{j\tilde{j}}\sqrt{\Sigma_j\Sigma_{\tilde{j}}}D_jD_{\tilde{j}}\right\}\right)
\end{align}
\subsection{Testing Procedure}
Reconsidering the sketched research question from above, it is necessary to be clear about the hypothesis that is supposed to be tested. Assume a single observation index $k^{-1} x_i^\top\bm{v}$, and we assume it to be $I_0$. The $H_0$-hypothesis is therefore that the industry-wide index $I$ equals $I_0$. For that purpose, corporation $i$ will be excluded from the sample. In order to be clear about the exclusion of $i$, the sample estimates are indexed by ${}_{-i}$, and a sample consists now of $n-1$ observations.\\
Since $\mathbb{V}[\hat{I}_{-i}]$ is not known, it has to be estimated. In the linear case, the estimate $\hat{\mathbb{V}}[\hat{I}_{-i}]$ is being obtained by using plug-in estimators for $\hat{\sigma}_{-ij}$ and $\hat{\rho}_{-ij\tilde{j}}$ (or, perhaps more conveniently, $\hat{\sigma}_{j\tilde{j}}$) for $j, \tilde{j}=1,2,..,k$, $j \neq \tilde{j}$. In addition, $D_j$ has to be evaluated at $\hat{S}_j$ in the nonlinear case (Raykov \& Marcoulides, 2004). The resulting test statistic is
\begin{equation}
T=\frac{\hat{I}_{-i}-I_0}{\sqrt{\hat{\mathbb{V}}[\hat{I}_{-i}]}}
\end{equation}
which has a Student's t-distribution with $n-k-1$ degrees of freedom. In dependence of the significance level of choice, a test decision can be made. Of course, this serves only as an example. Testing of one-sided hypotheses is also possible.\\
The second research question is whether two distinct industries differ significantly from each other. As an example, one could test whether the adoption of artificial intelligence and machine learning is different between the financial and marketing industry. The $H_0$-hypothesis is therefore $I' =I''$, where $I'$ corresponds to the Integrated Technology Adoption Index of the financial industry, and $I''$ corresponds to the same for the marketing industry, respectively (the examination concerning inequalities in the hypothesis is again possible but not covered). Let the samples consist of $n'$, and $n''$ observations, respectively, from which $\hat{I}'$ and $\hat{I}''$ are being estimated. The resulting test statistic (under the assumption of variance heterogeneity) is
\begin{equation}
T=\frac{\hat{I}'-\hat{I}''}{\sqrt{\hat{\mathbb{V}}[\hat{I}'] + \hat{\mathbb{V}}[\hat{I}'']}}
\end{equation}
It follows a Student's t-distribution with $\nu$ degrees of freedom, whereby $\nu$ is approximated by the Welch–Satterthwaite equation: $\nu\approx \frac{\left(\hat{\mathbb{V}}[\hat{I}']+\hat{\mathbb{V}}[\hat{I}'']\right)^2}{\hat{\mathbb{V}}[\hat{I}']^2/(n'-k)+\hat{\mathbb{V}}[\hat{I}'']^2/(n''-k)}$ (Welch, 1947).
\section{Application to an Integrated Technology Model for Blockchain-driven Business Innovations}
In this section, the developed framework is being applied to the Integrated Technology Adoption Model for Blockchain-driven Business Innovations by (Drljevic et al., 2022). First, the model is being introduced and interpreted in a way that it is tractable for the quantitative approach formulated above. Second, the framework is being applied for illustration purposes.
\subsection{The Model}
In Drljevic et al. (2022), the authors construct a novel, integrated technology adoption model with the objective of examining Blockchain-driven business innovations. Hereby, they focus on two main determinants in the process of technology adoption: novelty and complexity. In order to be able to apply well-established technology adoption models, they translate these determinants into acceptance and maturity. The authors describe acceptance as "a decision-based process, which is determined by user attitudes, values, and users’ intention to use blockchain technology", and maturity "as various stages of evolution of new technology adoption through different stages, which users move through in complex environments". In order to capture these two factors, their integrated model consists of the Technology Acceptance Model (TAM) and the Capability Maturity Model (CMM). The authors then conduct a questionnaire-based study to apply this novel approach empirically.\\
While the CMM consists of clearly sequential and ordered steps, each reflecting a evolutionary stage in the technology adoption process, the same does not exactly hold for TAM. Nonetheless, the relationships among the factors (external variables excluded) can be reduced to the following ordering: perceived ease of use - perceived usefulness - attitude towards usage - behavioral intention to use - actual system use (see Davis et al., 1989). The ordered states in the CMM are: initial - repeatable - defined - managed - optimizing (Paulk et al., 1991). As described above, we will add a stage to each model where there was no progress towards an adoption resulting in six stages for each.
\subsection{Quantitative Examination}
Considering the notation from above, it is clear that $k=2$. Since the models have clear abbreviations, the variables will be indexed by $j\in\{TAM,CMM\}$ instead of integer values. The two variables measured by questionnaire surveys are $X_{TAM}=\{a_{TAM}, p_{TAM}; a_{TAM}=0,1,2,3, 4, 5\}$ and $X_{CMM}=\{a_{CMM}, p_{CMM}; a_{CMM}=0,1,2,3,4, 5\}$. Thus, $m_{TAM}=5$ and $m_{CMM}=5$. Once observations are gathered, the sample scores can be estimated.
\begin{equation*}
\begin{aligned}
\hat{S}_{TAM}&=n^{-1}\sum_{i=1}^n x_{i,TAM}\\
\hat{S}_{CMM}&=n^{-1}\sum_{i=1}^n x_{i,CMM}
\end{aligned}
\end{equation*}
The concentration lies again on the linear index.
\begin{equation*}
\begin{aligned}
\hat{I}_{TAM}&=\frac{\hat{S}_{TAM}}{5}\\
\hat{I}_{CMM}&=\frac{\hat{S}_{CMM}}{5}
\end{aligned}
\end{equation*}
Finally, the estimate for the global technology adoption index is obtained.
\begin{equation*}
\hat{I}=\frac{\hat{I}_{TAM}+\hat{I}_{CMM}}{2}
\end{equation*}
\begin{figure}[H]
\centering
\includegraphics[scale=0.4]{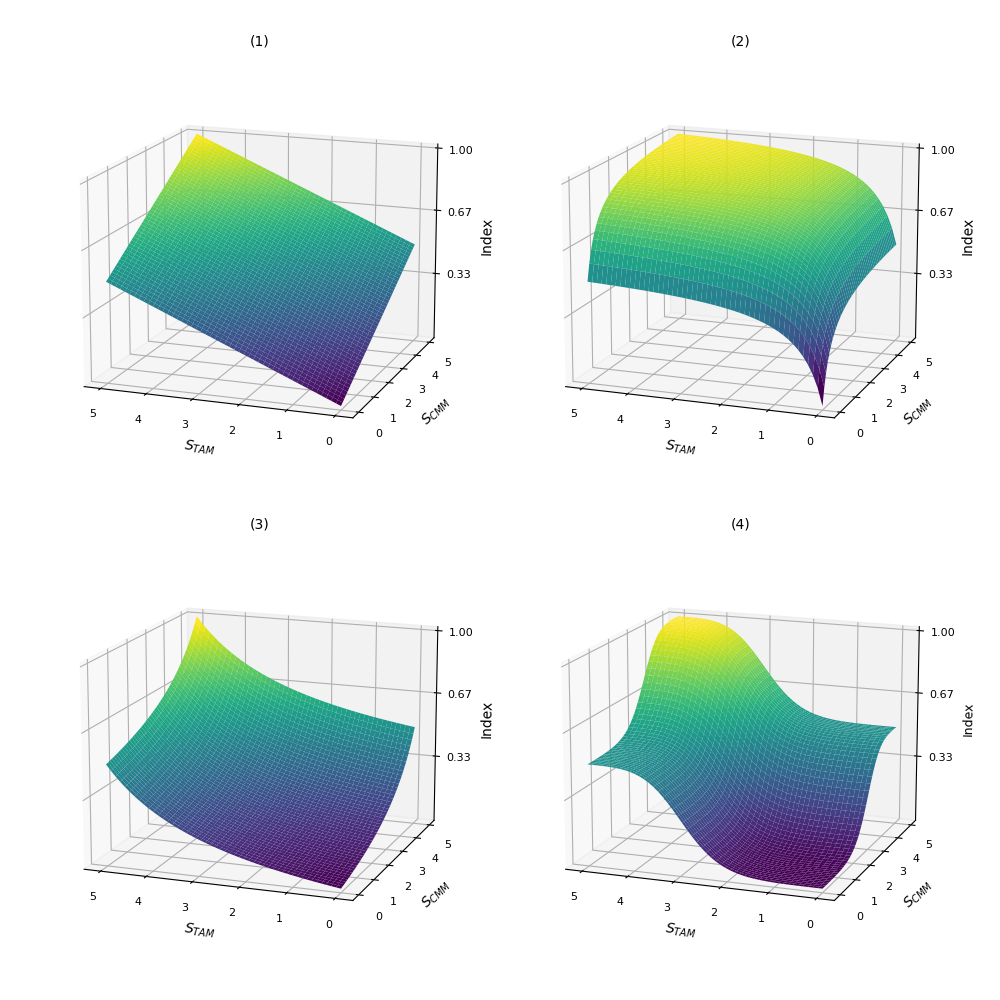}
\caption{Plots of the surface of the global index where each model has the same parameterization. (1): linear parameterization as in the example above. (2): concave parameterization. (3): convex parameterization. (4): s-shaped parameterization. With coordinates \{0,0\} the index is zero meaning no technology adoption at all. With coordinates \{5,5\} the index is 1 implying 100\% of technology adoption. However, it would implicate that all corporations are same in terms of the adoption level. We do not consider these two cases for testing.}
\end{figure}
As stated above, $\hat{I}$ is asymptotically normally distributed with
\begin{equation*}
\begin{aligned}
\mathbb{E}[\hat{I}]&=I\\
\mathbb{V}[\hat{I}]&=\frac{1}{5}\left(\Sigma_{TAM}+\Sigma_{CMM} + 2\rho_{TAM,CMM}\sqrt{\Sigma_{TAM}\Sigma_{CMM}}\right)\\
&=\frac{1}{5}\left(\Sigma_{TAM}+\Sigma_{CMM} + 2\Sigma_{TAM,CMM}\right)\\
&=\frac{1}{100n}\left(\sigma^2_{TAM}+\sigma^2_{CMM}+ 2n \sigma_{TAM,CMM}\right)
\end{aligned}
\end{equation*}
whereby $\Sigma_{TAM,CMM}\equiv\Cov(\hat{I}_{TAM},\hat{I}_{CMM})$ and $\sigma_{TAM,CMM}\equiv\Cov(\hat{S}_{TAM},\hat{S}_{CMM})$, respectively. At this point, we would like to mention that since TAM and CMM are meant to represent novelty and complexity, it could be assumed that these two are independent of each other. In this case $\sigma_{TAM, CMM}=0$ and the variance is reduced to $\mathbb{V}\left[\hat{I}\right]=\frac{\sigma^2_{TAM}+\sigma^2_{CMM}}{100n}$. Subsequently, a statistical test as outlined above can be conducted after estimating $\hat{\mathbb{V}}[\hat{I}]$.
\section{Outlook}
Above, we chose the most simple example to illustrate the proposed procedure. Both sub-indices are identical in their parameterization and, moreover, linear. A more realistic approach would be that the sub-indices are heterogeneously parameterized since the internal meaning of those two models is different. Illustrations of the global index surface with other parameter choices can be found in the figure below.
\begin{figure}[H]
\centering
\includegraphics[scale=0.33]{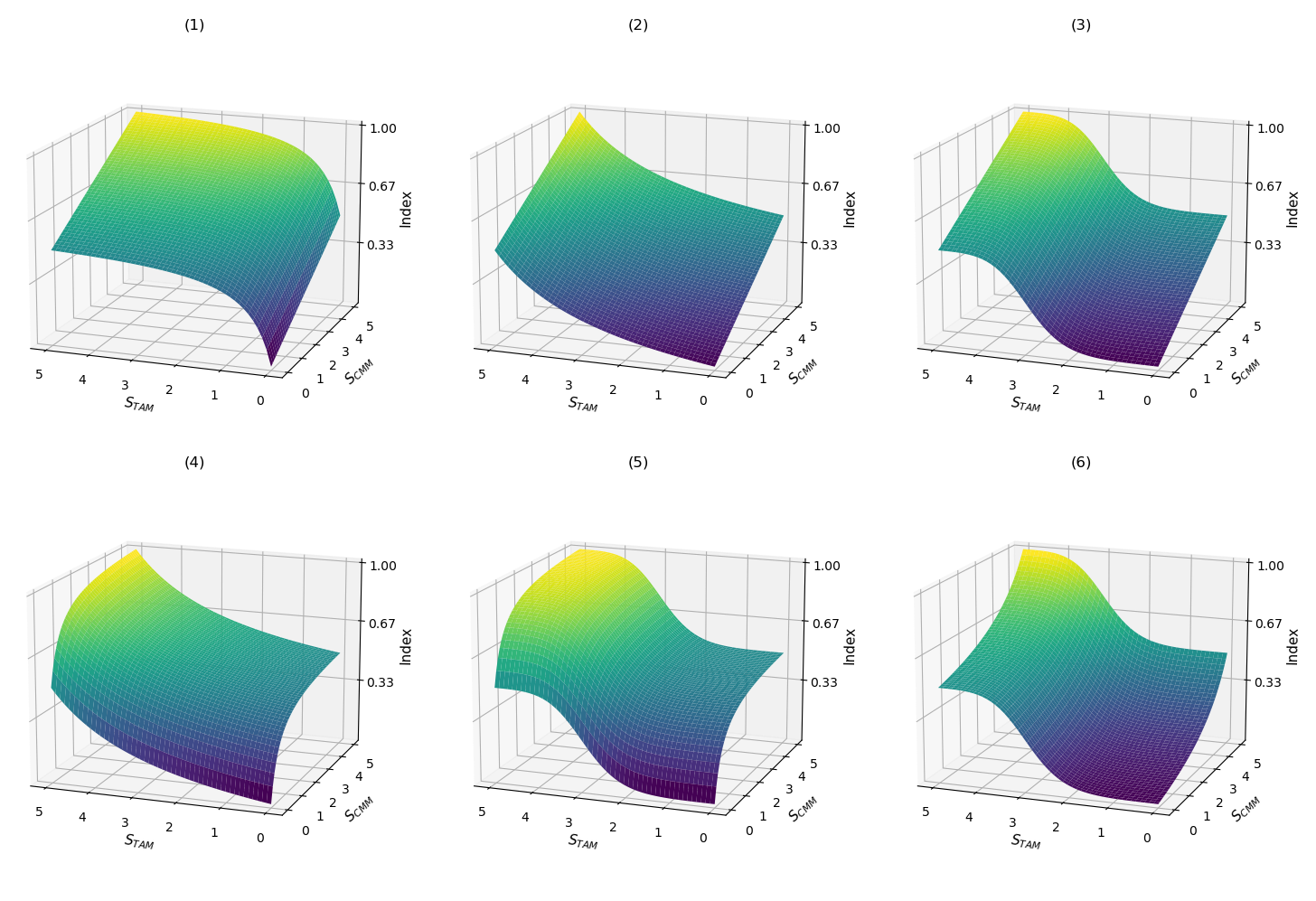}
\caption{Examples of the global index surface with heterogeneous parameterization of the sub-indices. (1): $I_{TAM}$ concave, $I_{CMM}$ linear. (2): $I_{TAM}$ convex, $I_{CMM}$ linear. (3): $I_{TAM}$ s-shaped, $I_{CMM}$ linear. (4): $I_{TAM}$ convex, $I_{CMM}$ concave. (5): $I_{TAM}$ s-shaped, $I_{CMM}$ concave. (6): $I_{TAM}$ s-shaped, $I_{CMM}$ convex.}
\end{figure}
Furthermore, we would like to encourage other researchers to analyze the adoption of a technology of their choice by combining other, and possibly more, models. In those cases, the global index surface will exist in a higher-dimensional space. Even though this might be a disadvantage with regards to illustration purposes, it is supposed to account for the complex structures the process of technology adoption can consist of. Which models, and which individual parameterization to choose, will depend highly on the technology of interest. Researchers must therefore rely on domain knowledge and perhaps even heuristics. Due to the reason that this is an open question for many specific research problems, it could lead to new discussions about the interpretation of existing models, and how technologies and innovations differ from each other. Since, to our best knowledge, the Integrated Technology Adoption Index is a novel tool in this field of research, we hope to strengthen a view on technology adoption that is integrated and accounts for its multi-dimensional nature. 
\section{Conclusion}
In this paper, the current state of technology adoption research was extended by a provision of a statistical framework enabling quantitative analyses. In particular, the Integrated Technology Adoption Index is introduced as a tool to make comparable and interpretable statements about the degree of technology adoption in any industry of choice. Furthermore, it is able to capture possible nonlinearities in the underlying models. Subsequently, the statistical properties and asymptotics are explored in order to form appropriate test statistics. This is supposed to give rise to the opportunity to examine various research questions that are subject to all kinds of technology that are exist or possibly developed in the future. Finally, the presented framework is applied to an integrated model of TAM and CMM. The corresponding solutions are derived.
\section*{References}
Ajzen, I. (1985). \textit{From intentions to actions: A theory of planned behavior} (pp. 11-39). Springer Berlin Heidelberg.\\\\
Ajzen, I. (1991). The theory of planned behavior. \textit{Organizational behavior and human decision processes}, 50(2), 179-211.\\\\
Ajzen, I., \& Fishbein, M. (1980). \textit{Understanding attitudes and predicting social behavior}. Englewood Cliffs, NJ: Prentice-Hall.Cox, 2012\\\\
Moore, G. C., \& Benbasat, I. (1991). Development of an instrument to measure the perceptions of adopting an information technology innovation. \textit{Information systems research}, 2(3), 192-222.\\\\
Chau, P. Y. (1996). An empirical assessment of a modified technology acceptance model. \textit{Journal of management information systems}, 13(2), 185-204.\\\\
Cox, Jr, L. A. (2012). Evaluating and improving risk formulas for allocating limited budgets to expensive risk‐reduction opportunities. \textit{Risk Analysis: An International Journal}, 32(7), 1244-1252.\\\\
Davis, F.D. (1985). \textit{A technology acceptance model for empirically testing new end-user information systems: theory and results}. Doctoral dissertation. MIT Sloan School of Management, Cambridge, MA\\\\
Davis, F. D., Bagozzi, R. P., \& Warshaw, P. R. (1989). User acceptance of computer technology: A comparison of two theoretical models. \textit{Management science}, 35(8), 982-1003.\\\\
Drljevic, N., Aranda, D. A., \& Stantchev, V. (2022). An Integrated Adoption Model to Manage Blockchain-Driven Business Innovation in a Sustainable Way. \textit{Sustainability}, 14(5), 2873.\\\\
Facchinetti, S., \& Osmetti, S. A. (2018). A risk index for ordinal variables and its statistical properties: A priority of intervention indicator in quality control framework. \textit{Quality and Reliability Engineering International}, 34(2), 265-275.\\\\
Fishbein, M., \& Ajzen, I. (1975). \textit{Belief, attitude, intention and behaviour: An introduction to theory and research}. Addison-Wesley.\\\\
Garzás, J., \& Paulk, M. C. (2013). A case study of software process improvement with CMMI‐DEV and Scrum in Spanish companies. \textit{Journal of Software: Evolution and Process}, 25(12), 1325-1333.\\\\
Härdle, W. K. \& Simar, L. (2015). \textit{Applied Multivariate Statistical Analysis}. Springer Berlin, Heidelberg\\\\
Herbsleb, J., Zubrow, D., Goldenson, D., Hayes, W., \& Paulk, M. (1997). Software quality and the capability maturity model. \textit{Communications of the ACM}, 40(6), 30-40.\\\\
Hu, P. J., Chau, P. Y., Sheng, O. R. L., \& Tam, K. Y. (1999). Examining the technology acceptance model using physician acceptance of telemedicine technology. \textit{Journal of management information systems}, 16(2), 91-112.\\\\
LaCaille, L. (2020). \textit{Theory of reasoned action. Encyclopedia of behavioral medicine}, 2231-2234.\\\\
Lee, Y. H., Hsieh, Y. C., \& Hsu, C. N. (2011). Adding innovation diffusion theory to the technology acceptance model: Supporting employees' intentions to use e-learning systems. \textit{Journal of Educational Technology \& Society}, 14(4), 124-137.\\\\
Lou, A. T. F. \& Li, E. Y. (2017). Integrating Innovation Diffusion Theory and the Technology Acceptance Model: The adoption of blockchain technology from business managers’ perspective. \textit{ICEB 2017}. Proceedings (Dubai, UAE). 44.\\\\
McGuire, E. G. (1996). Factors affecting the quality of software project management: An empirical study based on the Capability Maturity Model. \textit{Software Quality Journal}, 5, 305-317.\\\\
Paulk, M. C. (1997). The Capability Maturity Model for Software, version 2. \textit{Software Technology Conference}, Salt Lake City, April, 247-302.\\\\
Paulk, M. C., Curtis, B., \& Chrissis, M. B. (1991). \textit{Capability maturity model for software}. CARNEGIE-MELLON UNIV PITTSBURGH PA SOFTWARE ENGINEERING INST.\\\\
Paulk, M. C., Curtis, B., Chrissis, M. B., \& Weber, C. V. (1993). Capability maturity model, version 1.1. \textit{IEEE software}, 10(4), 18-27.\\\\
Paulk, M. C., S. M. Garcia, and M. B. Chrissis. (1996). The continuing improvement of the CMM, version 2. \textit{Fifth European Conference on Software Quality}, Dublin, September.\\\\
Poong, Y. S., \& Eze, U. C. (2008). TAM vs. PCI: An analysis on the theoretical model parsimony and robustness across cultures. \textit{Communications of the IBIMA}, 1(23), 198-201.\\\\
Raykov, T., \& Marcoulides, G. A. (2004). Using the delta method for approximate interval estimation of parameter functions in SEM. Structural Equation Modeling, 11(4), 621-637.\\\\
Rogers, E. M. (1962). \textit{Diffusion of innovations}. New York: Free Press.\\\\
Rogers, E. M. (2003). \textit{Diffusion of Innovations}. (5th ed.) New York: Free Press.\\\\
Sagnier, C., Loup-Escande, E., Lourdeaux, D., Thouvenin, I., \& Valléry, G. (2020). User acceptance of virtual reality: an extended technology acceptance model. \textit{International Journal of Human–Computer Interaction}, 36(11), 993-1007.\\\\
Szajna, B. (1996). Empirical evaluation of the revised technology acceptance model. \textit{Management science}, 42(1), 85-92.\\\\
Tomarchio, S. D., \& Punzo, A. (2019). Modelling the loss given default distribution via a family of zero‐and‐one inflated mixture models. \textit{Journal of the Royal Statistical Society Series A}, 182(4), 1247-1266.\\\\
Venkatesh, V., \& Davis, F. D. (2000). A theoretical extension of the technology acceptance model: Four longitudinal field studies. \textit{Management science}, 46(2), 186-204.\\\\
Venkatesh, V., \& Bala, H. (2008). Technology acceptance model 3 and a research agenda on interventions. \textit{Decision sciences}, 39(2), 273-315.\\\\
Vernon, R. (1966). International Investment and International Trade in the Product Cycle. \textit{The Quarterly Journal of Economics}, 80(2), 190–207.\\\\
Welch, B. L. (1947). The generalization of ‘STUDENT'S’problem when several different population varlances are involved. \textit{Biometrika}, 34(1-2), 28-35.
\end{document}